\documentclass[sigconf]{acmart}
\usepackage{CJKutf8}
\usepackage{graphicx}
\usepackage{natbib}
\usepackage{caption}
\usepackage{tabularx}
\usepackage{multirow}
\usepackage{subcaption}
\usepackage{xurl}
\usepackage{amsmath} 
\usepackage{xcolor}  
\usepackage{placeins}

\newcommand{\equalfirst}{\textsuperscript{\dag}}  
\newcommand{\corresp}{\textsuperscript{*}}        

\definecolor{deepgreen}{rgb}{0,0.6,0}

\AtBeginDocument{%
  }

\setcopyright{acmlicensed}
\copyrightyear{2018}
\acmYear{2018}
\acmDOI{XXXXXXX.XXXXXXX}
\acmConference[Conference acronym 'XX]{Make sure to enter the correct
  conference title from your rights confirmation email}{June 03--05,
  2018}{Woodstock, NY}
\acmISBN{978-1-4503-XXXX-X/2018/06}

\begin{document}

\title{Rethinking LLM Parametric Knowledge as Post-retrieval Confidence for Dynamic Retrieval and Reranking}

\thanks{\equalfirst These authors contributed equally to this work.}
\thanks{\corresp Corresponding authors.}





\author{Haoxiang Jin\equalfirst}
\affiliation{
  \institution{School of Computer Science and Technology, Xidian University}
  \city{Shaanxi}
  \country{China}
}
\email{jinhx@stu.xidian.edu.cn}

\author{Ronghan Li\equalfirst}
\affiliation{
  \institution{School of Computer Science and Technology, Xidian University}
  \city{Shaanxi}
  \country{China}
}
\email{lironghan@stu.xidian.edu.cn}

\author{Zixiang Lu\corresp}
\affiliation{
  \institution{School of Computer Science and Technology, Xidian University}
  \city{Shaanxi}
  \country{China}
}
\email{zxlu@xidian.edu.cn}

\author{Qiguang Miao\corresp}
\affiliation{
  \institution{School of Computer Science and Technology, Xidian University}
  \city{Shaanxi}
  \country{China}
}
\email{qgmiao@xidian.edu.cn}

\begin{abstract}
Large Language Models (LLMs) often generate inaccurate responses (hallucinations) when faced with questions beyond their knowledge scope. Retrieval-Augmented Generation (RAG) addresses this by leveraging external knowledge, but a critical challenge remains: determining whether retrieved contexts effectively enhance the model’s ability to answer specific queries. This challenge underscores the importance of knowledge boundary awareness, which current methods-relying on discrete labels or limited signals-fail to address adequately, as they overlook the rich information in LLMs’ continuous internal hidden states. To tackle this, we propose a novel post-retrieval knowledge filtering approach. First, we construct a confidence detection model based on LLMs’ internal hidden states to quantify how retrieved contexts enhance the model’s confidence. Using this model, we build a preference dataset (NQ\_Rerank) to fine-tune a reranker, enabling it to prioritize contexts preferred by the downstream LLM during reranking. Additionally, we introduce Confidence-Based Dynamic Retrieval (CBDR), which adaptively triggers retrieval based on the LLM’s initial confidence in the original question, reducing knowledge conflicts and improving efficiency. Experimental results demonstrate significant improvements in accuracy for context screening and end-to-end RAG performance, along with a notable reduction in retrieval costs while maintaining competitive accuracy.
\end{abstract}



\maketitle


\received{20 February 2007}
\received[revised]{12 March 2009}
\received[accepted]{5 June 2009}


\begin{figure*}[htbp]
    \centering
    \begin{subfigure}[t]{0.423\textwidth}
        \includegraphics[width=\linewidth]{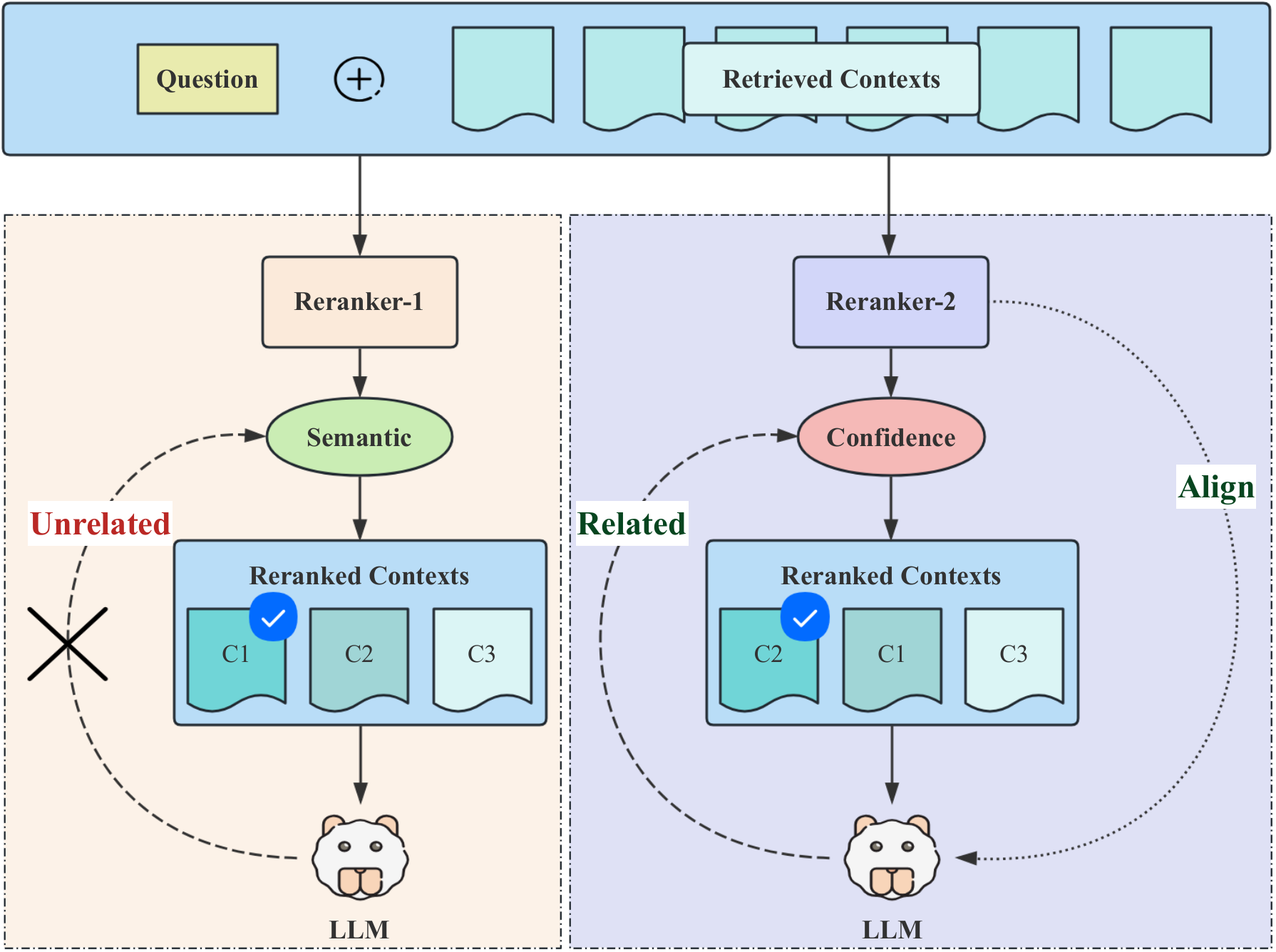}
        \caption{Comparison of Reranking Strategies}
        \label{FIG:1_a}
    \end{subfigure}
    \hfill 
    \begin{subfigure}[t]{0.57\textwidth}
        \includegraphics[width=\linewidth]{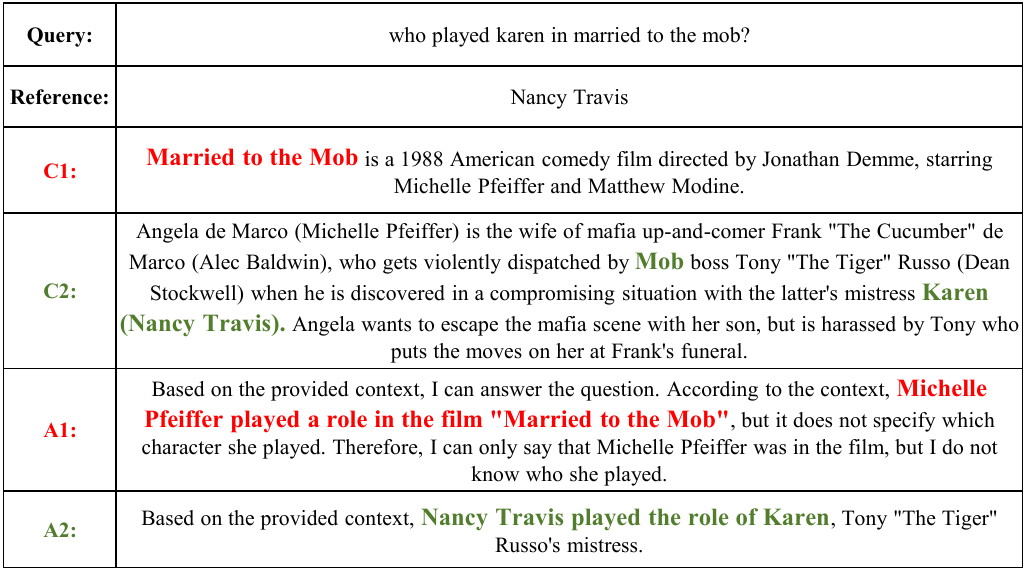}
        \caption{Performance Evaluation}
        \label{FIG:1_b}
    \end{subfigure}
    \caption{The left part of the diagram (Figure \ref{FIG:1_a}) contrasts two distinct reranking strategies for RAG systems. One employs a conventional similarity-based reranker, which prioritizes contexts solely through textual matching between questions and documents. The other leverages the LLM’s intrinsic preference, reranking contexts according to their ability to enhance the model’s answer confidence. The right part (Figure \ref{FIG:1_b}) provides a concrete example comparing the effectiveness of contexts reranked by Reranker-1 (similarity-based) and Reranker-2 (LLM-aligned). It demonstrates the responses generated by the same LLM when using each set of contexts, highlighting differences in answer quality, confidence, and relevance.}
    \label{FIG:1}
\end{figure*}

\section{Introduction}
Large Language Models (LLMs) have demonstrated exceptional performance in diverse text generation tasks, such as creative writing and text summarization \cite{zhao2023survey, touvron2023llama, hui2024qwen2}. However, when confronted with questions beyond their knowledge scope, they often generate plausible yet inaccurate responses—a phenomenon termed "hallucination" \cite{ji2023towards, martino2023knowledge}. To address questions unanswerable by model parameters alone, Retrieval-Augmented Generation (RAG) \cite{izacard2020leveraging, lewis2020retrieval} leverages external knowledge sources to expand the answerable question boundary . Yet this approach introduces a critical challenge : After retrieval, how can we precisely determine whether the acquired knowledge genuinely enhances the model’s ability to answer a specific query \cite{yoran2023making, fang2024enhancing, cuconasu2024power}?  The core of this challenge lies in effectively coordinating the model’s internal parametric knowledge with retrieved contexts to delineate the knowledge utilization scope of RAG system in post-retrieval stages.

This challenge underscores the importance of knowledge boundary awareness. Failures in answering questions typically stem from two causes: 1) Suboptimal prompt design, where inadequate prompting fails to unlock the model’s potential \cite{ji2023survey, dong2023statistical, yin2024benchmarking}. This can often be mitigated through prompt optimization strategies such as chain-of-thought \cite{wei2022chain} or self-verification prompts\cite{ren2023investigating} . 2) Fundamental unawareness of knowledge boundary, where the model cannot recognize its limitations and thus fails to abstain from answering beyond its competence \cite{yin2024benchmarking, li2024knowledge, zheng2025enhancing, ni2025towards}. This issue persists in RAG system: poor boundary awareness impedes judgment on the quality of retrieved contexts, which in turn makes it difficult to select useful ones that could enhance their reasoning on a given question  \cite{sun2025divide}. Enhancing such awareness is therefore crucial for improving RAG system accuracy.

Current research on knowledge boundary perception follows three primary paths: 1) Prompt-guided confidence estimation: Using engineered prompts to elicit self-assessed confidence scores \cite{yin2023large, ni2024llms, ni2024large}. 2) Multi-sample confidence aggregation: Estimating confidence via correctness rates across multiple responses to the same question \cite{brown2024large, longjohn2025statistical}. 3) Hidden-state-based confidence: Quantifying uncertainty through the model’s internal internal hidden states such as intermediate layer activations \cite{su2024unsupervised, chen2024inside, ni2025towards}. Most existing methods rely on discrete labels such as answerable and unanswerable, lacking analysis of how internal hidden states evolve before and after introducing external knowledge. Crucially, model`s internal hidden states, as continuous vector representations, encapsulate richer information than discrete tokens and may more faithfully reflect model confidence \cite{su2024unsupervised, chen2024inside, ni2025towards}.

We posit that LLM’s internal hidden state reflected confidence serves as a key indicator for evaluating whether retrieved contexts effectively enhances their question-answering capability. Furthermore, the confidence shift observed when LLM processes different retrieved contexts for the same question inherently reveals its preference among retrieved contexts. This intrinsic preference signal can guide reranker model to filter and optimize contexts post-retrieval, significantly boosting RAG system efficacy. As illustrated in Figure \ref{FIG:1_a}, the left side employs an unrelated Reranker + LLM to form a RAG system, which represents the most common current approach. On the right, building upon the left configuration, the Reranker aligns with the LLM’s intrinsic preference by ranking contexts that most enhance the LLM’s confidence in answering the question during the reranking process. Figure \ref{FIG:1_b} compares the helpfulness of C2 and C1 in assisting the same LLM in answering the question.

Inspired by work \cite{ni2025towards}, this paper proposes a novel method for post-retrieval knowledge filtering. First, we construct a confidence detection model based on the LLM's internal hidden states to quantify how much the retrieved contexts enhances the LLM's confidence. Leveraging this confidence detection model’s analysis, we build a preference dataset, NQ\_Rerank, which is then used to fine-tune a Reranker model. This fine-tuning enables the Reranker to prioritize contexts preferred by the downstream LLM during the reranking phase, thereby improving the accuracy of the RAG system in answer generation. Additionally, we introduce Confidence-Based Dynamic Retrieval(CBDR) that adaptively triggers the retrieval process based on the LLM’s initial confidence in the original question. This mechanism reduces the risk of knowledge conflicts while enhancing the RAG system’s overall efficiency . 

Experiments show our approach achieves: 1) 5.19\% improvement in post-retrieval contexts screening accuracy. 2) 4.70\% higher end-to-end RAG system accuracy. 3) 7.10\% reduction in retrieval costs (with dynamic retrieval enabled) while maintaining 5.60\% accuracy gains. This work establishes a quantifiable framework for identifying and extending knowledge boundary in RAG system and introduces a confidence-shift-based metric for evaluating retrieval augmentation effectiveness.

\section{Related Work}

\subsection{Knowledge Boundary of LLM}

Since the seminal study \cite{yin2024benchmarking} introduced the concept of knowledge boundaries to LLM research, it has become a cornerstone for evaluating model’s self-awareness. This work categorizes parametric knowledge into three types: \textbf{Prompt-agnostic Knowledge}, Correctly answerable regardless of query phrasing. \textbf{Prompt-sensitive Knowledge}, Answerable only under specific prompting strategies. \textbf{Unanswerable Knowledge}, Incapable of correct response under any prompt.

Current evaluation paradigms focus on measuring confidence levels regarding answerability:

\begin{description}
\item[Expression-based Confidence:] Leverages LLM’s instruction-following capability to elicit self-reported confidence via prompts \cite{ren2023investigating}.
\item[Sampling-based Confidence Estimation:] Uses multi-round sampling such as query paraphrasing or output variations to compute answer entropy and confidence scores \cite{feng2025improving}.
\item[Internal State-based Confidence Estimation:] Utilizes hidden states—particularly intermediate-layer activations when generating the first token—as continuous confidence indicators \cite{ni2025towards}.
\end{description}

Our work adopts the third approach \cite{ni2025towards}, using Mid\_Layer hidden states at the first response token generation as the confidence metric.

\subsection{Knowledge Boundary in RAG System}

The knowledge boundary in RAG system is typically defined as the knowledge space collectively formed by the LLM’s internal parametric knowledge and external retrieved knowledge. However, early evaluations of  RAG system capabilities predominantly focused on Retriever performance, overemphasizing the system’s reliance on external knowledge. As revealed by studies such as \cite{yoran2023making, fang2024enhancing, cuconasu2024power}, conflicts between externally retrieved knowledge and internal parametric knowledge may lead the model to produce low-confidence errors.

Consequently, subsequent research has shifted toward coordinating these dual knowledge sources and more precisely delineating RAG’s effective knowledge boundary. Approaches like \cite{marina2025llm} and \cite{yao2024seakr} analyze the internal states of  LLM to detect model uncertainty, dynamically determining whether to activate the retrieval process based on this characteristic. Most recently, the DTA \cite{sun2025divide} framework formally proposes the concept of the Knowledge Boundary of  RAG. This study categorizes potential queries into four quadrants based on the LLM’s inherent Parametric Knowledge Boundary $KB_p$ and the Retrieved Knowledge Boundary $KB_r$ provided by the Retriever, collectively delineating the holistic effective knowledge boundary of the RAG system.

\subsection{Preference Alignment in RAG system}

To enhance the efficiency of LLM in utilizing externally retrieved knowledge, it is essential to align the preferences between the Retriever and the LLMs within RAG system. Diverse studies have proposed distinct preference signals to guide this alignment: RE-PLUG \cite{shi2023replug} employs the probability of an LLM generating a correct answer as a preference signal to identify critical contexts; RRR \cite{cong2024query} utilizes the overall quality of LLM-generated responses as a preference metric; DPA-RAG \cite{dong2025understand} introduces a bidirectional alignment strategy to mitigate preference conflicts among different RAG components; RADIO \cite{jia2024bridging} constructs preference indicators based on the correctness of generated rationales, subsequently fine-tuning the reranker model to reconcile preference discrepancies between Retrievers and LLMs; SEAKR \cite{yao2024seakr} performs multi-round sampling for the same query and leverages the last-layer hidden states of LLMs at the end-of-sequence token (</s>) to compute a Gram matrix. This matrix quantifies model uncertainty, which serves as a preference signal for Reranker optimization.

The core innovation of this paper lies in proposing a novel preference metric: confidence shift, characterized by changes in the internal hidden states of LLM before and after exposure to different external knowledge. This metric is utilized to fine-tune the Reranker, enabling effective filtering of post-retrieval contexts. Compared to methods like SEAKR \cite{yao2024seakr} which similarly leverage internal hidden states but require multi-round sampling, but our confidence shift detection mechanism, based on a single forward pass, significantly reduces computational and temporal overhead. This efficiency is particularly advantageous for real-time application scenarios demanding low latency.

\begin{figure*}
	\centering
		\includegraphics[scale=.39]{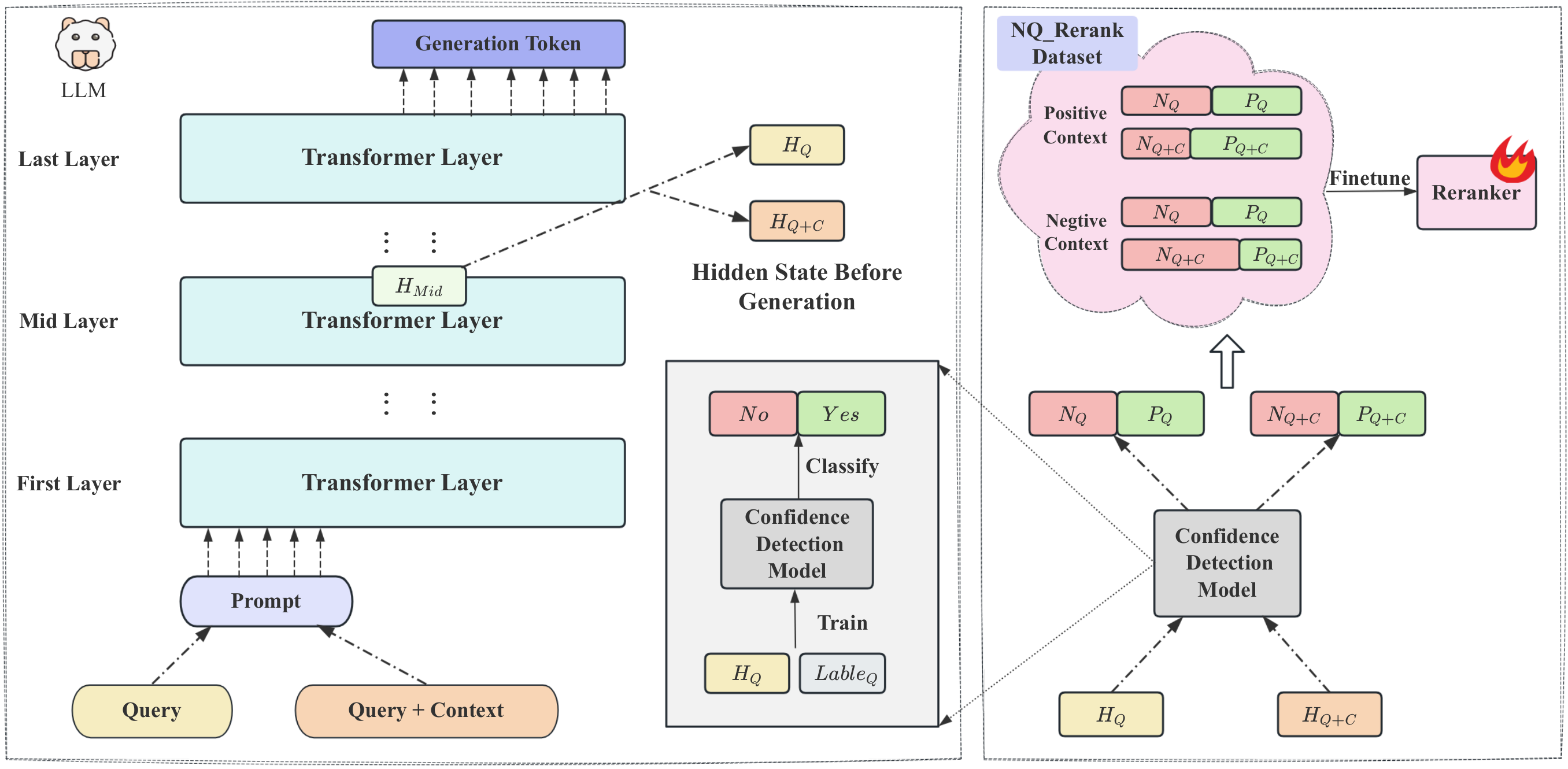}
	\caption{The complete process of aligning the Reranker with the target LLM involves constructing a preference dataset, NQ\_Rerank, by comparing the variations in the LLM's confidence when answering a Question under different contexts. This dataset is then used to fine-tune the Reranker model, aligning it with the target LLM's intrinsic preferences.}
	\label{FIG:2}
\end{figure*}

\section{Methods}

In this section, we provide a detailed exposition of the technical methodologies employed to assess the confidence of LLM in their responses to questions by leveraging their internal hidden states. We elaborate on the construction of a preference dataset based on variations in these internal hidden states and demonstrate how this dataset can be utilized to fine-tune a Reranker. Finally, we propose Confidence-Based Dynamic Retrieval(CBDR) that leverages the LLM’s self-confidence estimates to optimize information retrieval processes.

\begin{figure*}
	\centering
		\includegraphics[scale=.45]{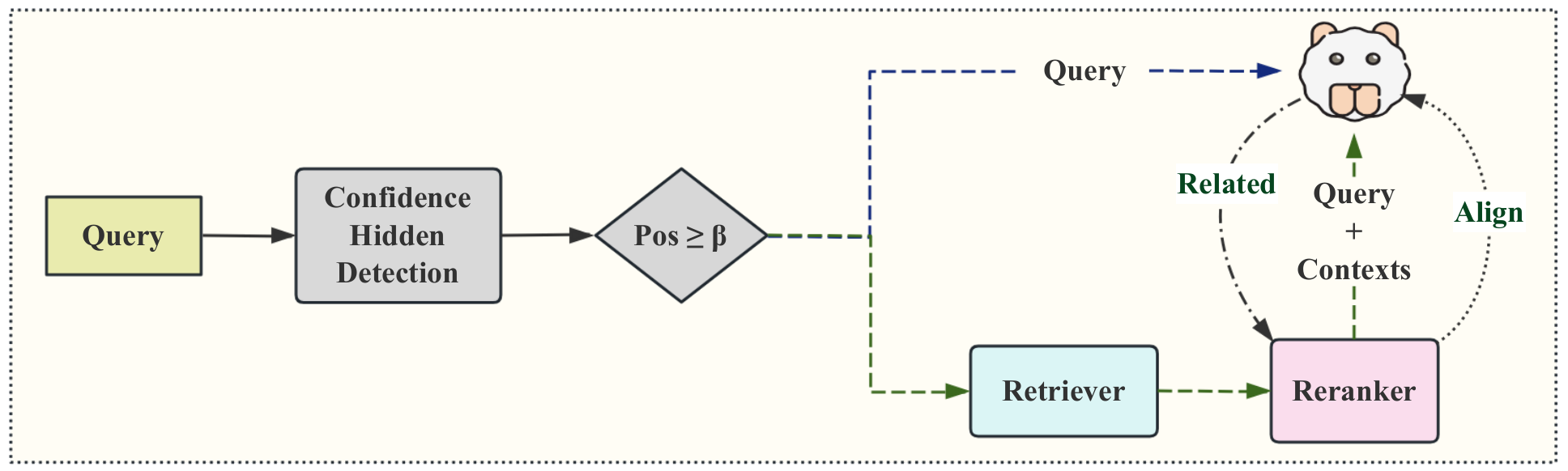}
	\caption{The workflow of Confidence-Based Dynamic Retrieval (CBDR). By varying the confidence threshold $\beta$, we balance the accuracy and retrieval cost of the Retrieval-Augmented Generation (RAG) system.}
	\label{FIG:3}
\end{figure*}

\subsection{Internal State Detection}

Human cognitive processes demonstrate that not all thought and reasoning rely on linguistic mediation; the explicit articulation of thought into language often entails a loss of information. This characteristic shares similarities with the operational mechanisms of modern LLM. LLM typically process input information through internal hidden layers and map complex latent representations into sequences of lexical tokens during the final output stage. This transformation, constrained by a fixed vocabulary, inevitably leads to partial information loss.

Recent studies, such as work \cite{skean2025layer} and work \cite{zhang2025reasoning}, reveal that the internal hidden states of LLMs contain richer information than their final outputs and exhibit stronger latent reasoning capabilities. Meanwhile, \cite{azaria2023internal} demonstrates that internal hidden states in internal hidden layers (specifically at Mid\_Layer which is Layer/2) effectively capture the model’s self-awareness. Furthermore, \cite{ni2025towards} indicates that LLMs can perceive their own knowledge boundaries prior to generating responses (Pre-Token).

\subsubsection{Confidence Estimation via Internal Hidden States}

Specifically, the workflow for self-confidence detection based on the internal hidden states of LLM is as follows: For a given target LLM M and a question Q, the model generates internal hidden state representations during inference, denoted as $H_{M,Q}$. Compared to the final token output, this state encapsulates more comprehensive information. Our confidence estimation process is defined as:

\begin{gather}
  C_{M,Q} = E(H_{M,Q}) \label{Eq:1} 
\end{gather}

As illustrated in left side of Figure \ref{FIG:2}, where E denotes the confidence detection model, and $C_{M,Q}$ is a binary classification label: $C_{M,Q} = 1$ indicates that LLM M is confident in correctly answering question q, whereas $C_{M,Q} = 0$ signifies that the LLM M perceives itself as incapable of responding accurately. Drawing on work \cite{ni2025towards} and related prior work, we select the internal hidden state vector at Mid\_Layer (Layer/2) of LLM M before generating the first answer token (Pre-Token) as $H_{M,Q}$.

The training data for model E is obtained by guiding LLM M to process questions from the NQ dataset \cite{47761}. We collect the internal hidden state $H_{M,Q}$ during inference and determine the correctness of the LLM M’s response based on the ground-truth answer to question Q, thereby constructing binary training samples ($H_{M,Q}$, $Label_Q$). Here, $Label_Q = 1$ indicates that the model answers question Q correctly, while $Label_Q = 0$ denotes an incorrect response. The training methodology for model E follows the approach described in work \cite{ni2025towards}. We performed data cleaning on the training dataset NQ \cite{47761} and analyzed the impact of different prompt designs on model reasoning.

\subsection{Preference Dataset}

\subsubsection{Preference Definition} \label{RD}

This study focuses on the post-retrieval processing stage within RAG system, with the aim of exploring how to rerank the retrieved contexts to maximize RAG system`s utility in enhancing the answer reasoning capabilities of downstream LLM.

Conventional Reranker are typically trained on datasets constructed based on semantic similarity between a question and contexts, and compute relevance scores by capturing complex semantic interactions through interactive encoding. While such general-purpose methods ensure model transferability and compatibility with diverse LLMs, they often fail to adequately incorporate the preferences of specific downstream LLM, thereby limiting the full potential of RAG system.

\begin{equation}
    \begin{aligned}
      Conf(H_{M,Q}) &= P(Label = 1 \mid E(H_{M,Q})) \\ &= Softmax(Z_1) \\ &= \frac{e^{z_1}}{e^{z_0} + e^{z_1}} \label{Eq:2} 
    \end{aligned}
\end{equation}

As illustrated in Figure \ref{FIG:2}, this paper defines the following preference criterion: a context $C$ is considered to exhibit a positive preference for the target LLM $M$ in answering question $Q$ if and only if it provides effective informational enhancement, satisfying the condition $Conf(H_{M,Q+C}) > Conf(H_{M,Q})$. Conversely, if it leads to a decrease in LLM`s confidence $Conf(H_{M,Q+C}) < Conf(H_{M,Q})$, the context $C$ is regarded as having a negative preference. As shown in Equation \ref{Eq:2}, the output of the Conf(-) function is defined as the probability of the $Label = 1$ assigned by model E. A softmax layer is appended to the final layer of model E to produce this probabilistic output.

\subsubsection{Dataset Construction}

We preprocess the NQ dataset \cite{47761} to obtain a series of $(Query, Contexts)$ tuple samples. For each sample, we record the internal hidden state at Mid\_Layer when the target LLM M generates its first token under the following two scenarios: 1) The state $H_{M,Q}$ when only the query $Q$ is provided; 2) The state $H_{M,Q+C_i}$ when both the query $Q$ and a context $C_i$ are provided (Where i iterates over the Contexts).

This yields a sequence of internal hidden states: 

\begin{align}
    [H_{M,Q},\ H_{M,Q+C_1},\ H_{M,Q+C_2},\ ...,\ H_{M,Q+C_i}] \notag
\end{align}

This sequence of states is then fed into the confidence hidden detection model E to obtain the probability value for the $Label = 1$ output by the softmax layer, resulting in a probability sequence:

\begin{align}
    [Conf(H_{M,Q}),\ Conf(H_{M,Q+C_1}),\ ...,\ Conf(H_{M,Q+C_i})] \notag
\end{align}

The enhancement effect of each context $C_i$ on LLM M’s response to question $Q$ is determined by comparing the change in model confidence after incorporating the context $C_i$:

\begin{align}
    Inc(Q,C_i) = Conf(H_{M,Q+C_i}) - Conf(H_{M,Q})
    \label{Eq:3}
\end{align}

If $Inc(Q,C_i) > 0$, the sample is labeled as a positive preference sample. If $Inc(Q,C_i) < 0$, it is labeled as a negative preference sample.

For each $(Query, Contexts)$ sample, all context $C_i$ are ranked according to $Inc(Q,C_i)$. The Top-K(K = 5) contexts with the highest increase are selected as positive examples, and the Top-K with the largest decrease are taken as negative examples. As illustrated in right side of Figure \ref{FIG:2}, this process constructs the final preference dataset, denoted as NQ\_Rerank.

\subsection{Reranker Fine-tuning}

To enhance the ability of the Reranker to identify the utility of contexts for the target LLM, we performed supervised fine-tuning on a base Reranker using the constructed preference dataset NQ\_Rerank. This fine-tuning process aims to achieve effective alignment between the Reranker and the target LLM’s preferences, enabling it to more accurately select contexts that significantly enhance the target LLM’s confidence in answering a given question $Q$.

During fine-tuning, the InfoNCE (Noise Contrastive Estimation) loss function was employed as the optimization objective:

\begin{gather}
    f(Q,C) = exp(\phi(Q,C)/\tau) \label{Eq:4} \\
    L = -log{\frac{f(Q,C^+)}{f(Q,C^+) + \sum_{i=1}^Nf(Q,C_i^-)}} \label{Eq:5}
\end{gather}

Where: $f(Q, C)$ denotes the relevance score between question $Q$ and context $C$ computed by the Reranker; $C^+$ represents the positive context; $C^-$ denotes the negative context; $\tau$ is the temperature parameter.

This loss function forces the model to increase the score margin between the positive context $C^+$ and a set of negative contexts $\{C^-\}$, thereby learning a ranking criterion consistent with the target LLM’s preferences.

\subsection{Confidence-Based Dynamic Retrieval}

Although the fine-tuned Reranker has achieved considerable alignment with the target LLM’s preferences and effectively improved the ranking priority of beneficial contexts, it still exhibits the following limitations:

\begin{description}
    \item[Risk of Irrelevant Context Interference:] The context set returned by the Retriever may contain few or no beneficial contexts. Under such circumstances, even after reranking, the Top-K results output by the Reranker may still include irrelevant or misleading context, which can interfere with the reasoning process of the downstream LLM and even lead to hallucinations.
    \item[Redundant Computational Overhead:] Typical Reranker often contain hundreds of millions to billions of parameters. Although single inference is relatively efficient, the cumulative computational cost of executing a full retrieval-reranking process for every query can be substantial, especially when the retrieved contexts have low relevance, resulting in significantly diminished cost-effectiveness.
\end{description}

To mitigate these issues and enhance the efficiency and reliability of the RAG system, we propose CBDR. The workflow of this strategy is illustrated in Figure \ref{FIG:3}: 1) If the target LLM exhibits high confidence in responding to the current query $Q$ that $Conf(H_{M,Q}) > \beta$, where $\beta$ is a predefined threshold, the retrieval and reranking steps are skipped, and the LLM generates the answer directly. 2) If the confidence score falls below the threshold $Conf(H_{M,Q}) < \beta$, the full retrieval process is initiated: the Retriever fetches a set of contexts, which are reranked by the fine-tuned Reranker, and the Top-K contexts are fed into the LLM along with the query for reasoning.

This strategy aims to maintain answer quality whenever possible while significantly reducing redundant computation for high-confidence questions and avoiding potential interference from low-quality retrieval results for known questions. The effectiveness of this strategy will be thoroughly validated in Section \ref{DRE} of the experiments.

\section{Experiments}

\begin{table*}[!ht]
    \centering
    \renewcommand{\arraystretch}{1.5}
    \setlength{\tabcolsep}{3pt}
    \caption{Performance Comparison of Different Rerankers on the NQ\_Rerank Test Set.}
    \label{tab:table_1}
    \resizebox{2\columnwidth}{!}{
\begin{tabular}{ccccccccccc}
\toprule
Reranker                                & Params & \multicolumn{3}{c}{Top-1}                        & \multicolumn{3}{c}{Top-3}                        & \multicolumn{3}{c}{Top-5}                        \\
                                        &        & Precision      & Recall         & MRR            & Precision      & Recall         & MRR            & Precision      & Recall         & MRR            \\
\midrule
gte\_passage-ranking\_multilingual-base & 304M   & 85.52          & 29.47          & 85.52          & 71.45          & 62.66          & 90.37          & 60.98          & 82.53          & 90.99          \\
Qwen3-Reranker                          & 4B     & 81.74          & 27.62          & 81.74          & 70.92          & 62.33          & 88.15          & 61.71          & 83.53          & 88.93          \\
Qwen3-Reranker                          & 8B     & \underline {87.25}    & \underline {30.47}    & \underline {87.25}    & \underline {74.35}    & \underline {65.15}    & \underline {91.65}    & \underline {64.22}    & \underline {86.42}    & \underline {92.19}    \\
bge-reranker-v2-m3                      & 568M   & 86.01          & 29.45          & 86.01          & 72.62          & 63.61          & 90.47          & 62.40          & 84.01          & 91.07          \\
bge-reranker-v2-m3-ft (Ours)             & 568M   & \textbf{91.20} & \textbf{32.01} & \textbf{91.20} & \textbf{76.98} & \textbf{67.14} & \textbf{94.40} & \textbf{65.64} & \textbf{87.97} & \textbf{94.72} \\
\bottomrule
\end{tabular}
}
\end{table*}

This section aims to systematically evaluate the effectiveness of the proposed method, focusing on the following three aspects: 1) We assess the impact of preference-aligned fine-tuning on the performance of the Reranker by examining the improvement in ranking tasks after fine-tuning with the constructed dataset NQ\_Rerank. 2) We evaluate the influence of the preference-aligned Reranker on the RAG system, analyzing whether the internal hidden states of the LLM can reliably reflect its preference for retrieved contexts. 3) We examine the effect of the CBDR on the efficiency and accuracy of the RAG system, verifying the performance and advantages of the confidence-based retrieval scheduling strategy in practical applications.

It should be noted that, since this study employs the Reranker to align with the preferences of downstream LLM and designs experiments accordingly to evaluate the improvements, the Retriever module has been intentionally excluded from the experimental setup. Instead, all Rerankers are provided with the same set of retrieved contexts to ensure a fair evaluation of their reranking performance under consistent contextual inputs.

\subsection{Experimental Setup}

\subsubsection{Datasets}

We take two representative open-domain QA benchmark datasets, including the Natural Questions (NQ) dataset \cite{47761} and the HotpotQA dataset \cite{yang2018hotpotqa}. The NQ dataset is constructed from real Google search questions and contains retrieved relevant contexts along with human-annotated short and long answers. HotpotQA is a question-answering dataset comprising examples that require multi-step reasoning.

All data used for training in this study were derived from the NQ dataset. The specific data partitions are described below:

\begin{description}
    \item[Dataset for Hidden Detection Model E:] The training set \\ (Train) consists of 1,000 positive samples (questions answered correctly by the target LLM M) and 1,000 negative samples (questions answered incorrectly by the target LLM M), selected from the NQ dataset; the development set (Dev) contains 300 positive and 300 negative samples and was used for hyperparameter tuning and early stopping; the test set (Test) includes 500 positive and 500 negative samples for the final performance evaluation of the mode.
    \item[The preference dataset NQ\_Rerank:] Constructed based on the NQ\_Retrieval\footnote{https://modelscope.cn/datasets/sentence-transformers/NQ-retrieval} dataset—a structured version of the NQ dataset adapted for retrieval tasks. The final version of the dataset excluded data items lacking valid positive-context or negative-context examples. The resulting dataset comprises 7,622 training samples and 1,216 evaluation samples.
\end{description}

\subsubsection{LLMs}

We conducted experiments on two representative open-source models, including Llama3-8B-Instruct \cite{dubey2024llama} and Qwen2.5-7B-Instruct \cite{qwen2.5}. The experiments uniformly employed Llama3-8B-Instruct as the base LLM for downstream task reasoning. During inference, the temperature of the model was set to 1.0, and tokens were selected using a greedy decoding strategy.

\subsubsection{Rerankers}

For a comprehensive comparison, we selected four representative Rerankers as backbone models and conducted preference alignment fine-tuning experiments specifically on bge-reranker-v2-m3:

\begin{description}
    \item[gte\_passage-ranking\_multilingual-base: ] A multilingual \\Reranker proposed by Alibaba DAMO Academy. Public evaluations indicate that it outperforms other models of similar scale in multilingual retrieval tasks.
    \item[Qwen3-Reranker-4B: ] A 4-billion-parameter version based on the Qwen foundation model, specifically optimized for text embedding and reranking tasks.
    \item[Qwen3-Reranker-8B: ] A 8-billion-parameter version based on the Qwen foundation model, Qwen3-Reranker-8B represents one of the state-of-the-art publicly available rerank models.
    \item[bge-reranker-v2-m3: ] A lightweight reranker known for its strong multilingual support and efficient inference speed.
    \item[bge-reranker-v2-m3-ft(Ours): ] To validate the effectiveness of aligning with the target LLM’s preferences via confidence estimation, we conducted supervised fine-tuning on the bge-reranker-v2-m3 model using the constructed NQ\_Rerank preference dataset.
\end{description}

\subsubsection{Evaluation Metrics}

\begin{equation}
    MRR@K = \frac{1}{|Q|}\sum_{i=1}^{|Q|}\frac{1}{rank_i}
\end{equation}

Following mainstream evaluation practices for Rerankers , we adopt the following metrics to assess performance: 1) Precision@K measures the proportion of effectively positive contexts among the Top-K ranked results. 2) Recall@K evaluates the ratio of successfully ranked positive contexts within the Top-K results relative to all relevant contexts. 3) Mean Reciprocal Rank (MRR@K) represents the average reciprocal rank of the first relevant context across all questions. Where $rank_i$ denotes the position of the first relevant context for the i-th query (positions beyond K are excluded from calculation).

Given that the number of positive contexts or negative contexts per query item in the NQ\_Rerank dataset ranges from [1, 5], we select Top-k values of k={1, 3, 5} for evaluation to align with the data characteristics.

\subsubsection{Implementation Details}

During the training of the internal hidden detection model E, the initial learning rate was set to $5e^{-5}$, the dropout rate was configured to 0.5, and the training was conducted over 30 epochs.

For the fine-tuning of the bge-reranker-v2-m3 model, the initial learning rate was set to $6e^{-5}$, weight decay was configured to 0.01, the maximum query length (query\_max\_len) was set to 128, the maximum passage length (passage\_max\_len) was set to 512, and the training was performed for 1 epoch.

\subsection{Results}

\begin{table*}[!ht]
    \centering
    \renewcommand{\arraystretch}{1.5}
    \setlength{\tabcolsep}{3pt}
    \caption{Accuracy of RAG Systems with Different Reranker and LLM Combinations.}
    \label{tab:table_2}
    \resizebox{2\columnwidth}{!}{
\begin{tabular}{clcclcclcc}
\toprule
LLM                 &  & Reranker                                & Params &  & \multicolumn{2}{c}{HotpotQA} &  & \multicolumn{2}{c}{NQ} \\
                    &  &                                         &        &  & Top-1         & Top-3        &  & Top-1      & Top-3     \\
\midrule
                    &  & gte\_passage-ranking\_multilingual-base & 304M   &  & 47.20         & 51.80        &  & \underline {63.80}      & 67.60     \\
                    &  & Qwen3-Reranker                          & 4B     &  & 42.30         & 50.10        &  & 50.70      & 64.00     \\
Qwen2.5-7B-Instruct &  & Qwen3-Reranker                          & 8B     &  & \underline {47.50}         & \underline {51.90}        &  & 56.30      & 68.80     \\
                    &  & bge-reranker-v2-m3                      & 568M   &  & 47.20         & \textbf{53.30}        &  & \textbf{64.20}      & \underline {69.70}     \\
                    &  & bge-reranker-v2-m3-ft (Ours)            & 568M   &  & \textbf{48.70}         & \textbf{53.30}        &  & 63.40      & \textbf{69.90}     \\
\midrule
                    &  & gte\_passage-ranking\_multilingual-base & 304M   &  & \textbf{48.80}         & 50.20        &  & 60.10      & 60.70     \\
                    &  & Qwen3-Reranker                          & 4B     &  & 40.70         & 48.00        &  & 49.70      & 62.30     \\
Llama3-8B-Instruct  &  & Qwen3-Reranker                          & 8B     &  & \underline {48.40}         & 50.10        &  & 55.20      & \textbf{68.80}     \\
                    &  & bge-reranker-v2-m3                      & 568M   &  & 46.60         & \underline {51.40}        &  & \underline {61.50}      & 62.20     \\
                    &  & bge-reranker-v2-m3-ft (Ours)            & 568M   &  & 48.00 $\color{deepgreen}\uparrow{}$        & \textbf{52.20} $\color{deepgreen}\uparrow{}$       &  & \textbf{62.60} $\color{deepgreen}\uparrow{}$     & \underline {66.90} $\color{deepgreen}\uparrow{}$   \\
\bottomrule
\end{tabular}
}
\end{table*}

\begin{table}[!ht]
    \centering
    \renewcommand{\arraystretch}{1.5}
    \setlength{\tabcolsep}{3pt}
    \caption{Efficiency-Accuracy Trade-off of the Confidence-Based Dynamic Retrieval Strategy.}
    \label{tab:table_3}
\begin{tabular}{cccccc}
\toprule
Reranker                     &  & \multicolumn{4}{c}{NQ}                 \\
                             &  & Top-1 & Top-3 & $\mathrm{\beta}$    & RR $\color{deepgreen}\downarrow{}$ \\
\midrule
bge-reranker-v2-m3           &  & 61.50 & 62.20 & 0    & 100             \\
bge-reranker-v2-m3-ft (Ours) &  & \textbf{62.60} & \underline{66.90} & 0    & 100             \\
bge-reranker-v2-m3-ft (Ours) &  & \underline{62.40} & 66.10 & 0.95 & \textbf{83.30}            \\
bge-reranker-v2-m3-ft (Ours) &  & 61.70 & \textbf{67.80} & 0.98 & \underline{92.90}           \\
\bottomrule
\end{tabular}
\end{table}

\subsubsection{Reranker Performance}

To evaluate whether the fine-tuned Reranker bge-reranker-v2-m3-ft can more effectively select retrieved contexts suitable for the downstream LLM Llama3-8B-Instruct, comparative experiments were conducted on the NQ\_Rerank test set. The experimental setup was as follows: each Reranker was provided with a query and its corresponding set of contextual documents (including both positive contexts and negative contexts). Each Reranker then output a reranked list of Top-K relevant documents (where K $\in$ {1, 3, 5}). The performance of each Reranker was comprehensively evaluated using the Precision@K, Recall@K, and MRR@K metrics.

The experimental results, summarized in Table \ref{tab:table_1}, are as follows: 
\begin{description}
    \item[1.] The fine-tuned model bge-reranker-v2-m3-ft, which underwent preference alignment training using the NQ\_Rerank dataset, achieved optimal performance across all evaluation metrics (K $\in$ {1, 3, 5}).
    \item[2.] Notable improvements were observed particularly in Top-1 performance. Compared to the second-best model Qwen3\\\_Rerank\_8B, Precision@1 and MRR@1 increased by +3.95 percentage points (pp) while Recall@1 improved by +1.54 pp; when compared to the baseline model bge-reranker-v2-m3 before fine-tuning, Precision@1 and MRR@1 increased by +5.19 pp and Recall@1 improved by +2.56 pp.
\end{description}

\subsubsection{RAG System Accuracy}

To further investigate whether a fine-tuned Reranker can enhance the final performance of a Retrieval-Augmented Generation (RAG) system, we constructed multiple "Reranker + LLM" combined systems for comparative experiments. The experimental setup was as follows: each Reranker performed re-ranking on the query and its corresponding set of context documents, selected the Top-K (K $\in$ {1, 3}) documents, and fed them to the downstream LLM for answer generation. The overall system performance was ultimately evaluated by assessing the precision of the generated answers.

The experimental results, as shown in Table \ref{tab:table_2}, yielded the following main findings:

\begin{description}
    \item[1.] When the downstream LLM was Llama3-8B-Instruct, the RAG system utilizing the bge-reranker-v2-m3-ft as the Reranker consistently achieved higher accuracy than the system using the original bge-reranker-v2-m3, with a maximum improvement of up to +4.7 pp. Furthermore, this combination achieved optimal or near-optimal performance on both the NQ and HotpotQA datasets.
    \item[2.] When the downstream LLM was Qwen2.5-7B-Instruct, the RAG systems employing either the fine-tuned or the original Reranker demonstrated comparable performance in terms of accuracy, with no significant differences observed. This to some extent demonstrates the robustness of bge-reranker-v2-m3-ft.
\end{description}

\subsubsection{Dynamic Retrieval Efficiency} \label{DRE}

We also evaluated the impact of a dynamic retrieval strategy based on the confidence level of the downstream LLM on the performance of the RAG system. The experiment measured the system's accuracy (Precision) and the proportion of overhead saved due to skipped retrieval under different confidence thresholds, as summarized in Table \ref{tab:table_3}. The main findings are as follows:

\begin{description}
    \item[1. ] Enabling the dynamic retrieval mechanism allowed the RAG system to significantly reduce retrieval overhead by assessing the LLM's confidence in addressing the query. Furthermore, under the setting with a reranking scope of Top-3 contexts, the system's accuracy improved by 0.9 percentage points.
    \item[2. ] In the Top-1 scenario, the strategy with dynamic retrieval enabled resulted in a slightly lower accuracy—by 0.2 percentage points—compared to the strategy without dynamic retrieval.
\end{description}

\section{Discussion}

Based on the experimental results presented, we draw the following conclusions:

\subsection{Model Scale Positively Correlates with Reranker Performance}

A positive correlation is observed between model scale and performance among the four un-tuned base Rerankers. Generally, performance improves as model size and capability increase. Taking the Qwen3\_Reranker series as an example: although its absolute performance remains relatively low, Qwen3\_Rerank\_8B(8 billion parameters) significantly outperforms its lighter counterpart Qwen3\_Rerank\_4B(4 billion parameters) and achieves the best results among the base models. This trend demonstrates that 1) the preference dataset NQ\_Rerank, constructed based on LLM confidence, possesses intrinsic validity in distinguishing model capabilities; and 2) the proposed confidence-change-based preference definition (Section \ref{RD}) is well-founded.

\subsection{Preference-Aligned Fine-Tuning Yields Substantial Gains}

Preference-aligned fine-tuning leads to substantial performance gains. The fine-tuned Reranker bge-reranker-v2-m3-ft outperforms all baseline Rerankers across all evaluation metrics. This strongly indicates that supervised fine-tuning using the NQ\_Rerank dataset effectively aligns the Reranker model with the preferences of the target downstream LLM (Llama3-8B-Instruct), making it more inclined to select contexts that increase the LLM’s confidence in answering a given query. Significant improvements in Top-1 accuracy (Precision@1) and Mean Reciprocal Rank (MRR@1) further highlight the fine-tuned Reranker’s enhanced capability in high-precision retrieval of critical documents.

\subsection{Effectiveness of Preference Alignment Depends on the Downstream LLM}

The effectiveness of preference alignment varies depending on the downstream LLM. Results from Table \ref{tab:table_2} show that when using the same fine-tuned Reranker bge-reranker-v2-m3-ft, accuracy improvements reached up to 4.7 pp with Llama3-8B-Instruct, whereas negligible gains were observed with Qwen2.5-7B-Instruct. This suggests that the success of preference alignment is partially dependent on the architecture and training methodology of the target LLM, and that the preferences embedded during fine-tuning must be compatible with the specific LLM being used.

\subsection{Dynamic Retrieval Balances Performance and Efficiency}

The dynamic retrieval strategy effectively balances performance and efficiency. By incorporating CBDR, the system significantly reduces retrieval overhead while maintaining competitive accuracy—improving by 0.9 pp under the Top-3 setting. Although a slight decrease in accuracy (0.2 pp) was observed in the Top-1 scenario, this approach demonstrates practical potential for balancing efficiency and effectiveness in real-world applications.

\section{Conclusion}

This paper establishes internal hidden state confidence dynamics as a principled signal for optimizing RAG systems. By quantifying confidence shifts induced by retrieved contexts, we enable precise Reranker alignment and adaptive retrieval activation. Our framework CBDR has brought more efficient performance to the RAG system, which has practical application value. Our work bridges parametric and external knowledge while providing a generalizable paradigm for evaluating knowledge boundary interactions. Future work will extend this approach to multimodal RAG and multi-documents knowledge scenarios.



\bibliographystyle{unsrt} 
\bibliography{reference}

\appendix

\end{document}